\documentclass[a4paper,11pt]{article}
\usepackage{pos}

\title{De Sitter, gravitino mass and the swampland}

\author*[a]{Niccol\`o Cribiori}

\affiliation[a]{Max-Planck-Institut f\"ur Physik (Werner-Heisenberg-Institut),\\
F\"ohringer Ring 6, 80805 M\"unchen, Germany}

\emailAdd{cribiori@mpp.mpg.de}

\abstract{In this contribution, we review a general argument showing that de Sitter critical points of extended supergravity are in tension with the magnetic weak gravity conjecture if the gravitino mass is vanishing. Motivated by this assumption, we review then the gravitino conjecture, which states that the limit of vanishing gravitino mass is pathological for the effective field theory description. Finally, we discuss more in general the fate of de Sitter critical points (with massless gravitini) in supergravity and comment on extensions of these works along various directions. Part of the material here presented is unpublished.}

\FullConference{%
  Corfu Summer Institute 2021 "School and Workshops on Elementary Particle Physics and Gravity"\\
  29 August - 9 October 2021\\
  Corfu, Greece
}

\begin{document}
\maketitle

\section{Introduction}

Understanding the origin of the small, positive value of dark energy density measured at present \cite{Planck:2018vyg} is one of the most challenging open problems in fundamental physics. One could hope that splitting the problem into two parts might help in seeking for its solution, but this is not quite the case. Indeed, on the one side one should understand how to embed a positive cosmological constant, or more generally a positive background energy density, into a consistent UV complete theory of quantum gravity. On the other side, one should understand the precise details of the supersymmetry breaking mechanism, which necessarily occurs when the background energy is positive, but it can also occur more generally. Both of these problems are challenging in themselves. 

Effective field theories are the typical framework in which to perform such investigations. However, the presence of dynamical gravity in the setup might change the standard paradigm according to which these theories are constructed, based on a genuinely wilsonian approach. That this is indeed the case is the central idea underlying the swampland program  \cite{Palti:2019pca,vanBeest:2021lhn}, which collects a number of conjectures that are supposed to encode essential features of quantum gravity. Following a purely bottom-up perspective, one should thus assume swampland conjectures as actual consistency criteria (or perhaps even principles) of quantum gravity and check whether or not they are satisfied in any given effective theory.

At present, it is fair to say that not all of the swampland conjectures share the same level of rigor and accuracy. For example, the absence of global symmetries \cite{Banks:1988yz} or the weak gravity conjecture \cite{Arkani-Hamed:2006emk} are widely believed to be established facts of quantum gravity, as they have been tested extensively through the years; more recent conjectures, such that the one forbidding de Sitter vacua \cite{Obied:2018sgi} and its various refinements, are instead still under debate. Nevertheless, there is evidence that all of the swampland conjectures should be connected to one another to form some sort of web. This implies that, even if a precise statement on the fate of de Sitter vacua in quantum gravity might not yet be known, some piece of information should be present in others, possibly more established, conjectures.

This line of reasoning has been followed by \cite{Cribiori:2020use,DallAgata:2021nnr}, where it is shown that, under certain assumptions, de Sitter vacua in extended supergravity are in tension with the magnetic weak gravity conjecture. The assumptions are the presence of an unbroken abelian gauge group (needed to apply the weak gravity conjecture), and a vanishing gravitino mass on the vacuum.\footnote{As in \cite{Cribiori:2020use,DallAgata:2021nnr}, here we always mean Lagrangian mass.}  In those setups in which the argument of \cite{Cribiori:2020use,DallAgata:2021nnr} applies, any known version of the de Sitter conjecture seems thus to be redundant, since the fate of de Sitter critical points is already dictated by the weak gravity conjecture. This material is reviewed in section \ref{sec_WGCvsdS}.

Given that the interplay between the weak gravity conjecture and a vanishing gravitino mass has an impact on the fate of de Sitter critical points, one could wonder whether some piece of information about quantum gravity is actually encoded into the gravitino mass itself. Indeed, the gravitino is the superpartner of the graviton and it is thus expected to be related to (quantum) gravity. Furthermore, in a (quasi)-flat background, similar to the one we are living in at present, the gravitino mass is a good estimation of the supersymmetry breaking scale. Whether or not quantum gravity can say something non-trivial in its respect is thus of clear phenomenological interest. This motivated  \cite{Cribiori:2021gbf,Castellano:2021yye} to propose a new swampland conjecture, stating that the limit of vanishing gravitino mass corresponds to a breakdown of the effective description. This material is reviewed in section \ref{sec_gravconj}.

Finally, in section \ref{sec_fincomm} we comment on possible extensions of these works and ideas. We work in Planck units.

\section{Weak gravity versus de Sitter}
\label{sec_WGCvsdS}

In this section, we review the general argument of \cite{Cribiori:2020use,DallAgata:2021nnr} showing that de Sitter critical points of extended supergravity with vanishing gravitino mass and with an unbroken abelian factor are in tension with the weak gravity conjecture. As discussed at the end of the section, this excludes almost all known examples in the literature (notably, \cite{DallAgata:2012plb} found an unstable de Sitter vacuum with massive gravitini and thus the argument here reviewed does not apply).

\subsection{A weak gravity constraint on de Sitter}
\label{arg}

Any effective field theory is characterised by at least two energy scales, namely an IR cutoff, $\Lambda_{IR}$, and a UV cutoff, $\Lambda_{UV}$. It natural to ask for the existence of an hierarchy between them,
\begin{equation}
\label{natcond}
\Lambda_{IR} \ll \Lambda_{UV},
\end{equation}
otherwise there would be almost no room for the theory to live in. The question is then what these two scales are. 
In a de Sitter space, there is a natural notion of IR cutoff given by the Hubble scale
\begin{equation}
\label{LIRH}
\Lambda_{IR} \sim H,
\end{equation}
since a distance of order $1/H$ is the longest length that can be measured. The choice of $\Lambda_{UV}$ is instead more subtle. Assuming gravity to behave classically in the effective description, a natural guess would be $\Lambda_{UV} \sim M_p$. However, one of the main lessons of the swampland program is that gravity is not genuinely wilsonian and thus the UV cutoff of a given effective theory can be lowered from its naive expectation by (quantum) gravity effects. A realisation of this scenario is the magnetic weak gravity conjecture \cite{Arkani-Hamed:2006emk}, which states that in an effective theory with gravity and with an abelian gauge coupling $g$, the UV cutoff is bounded by
\begin{equation}
\label{mWGC}
\Lambda_{UV} \lesssim g M_p.
\end{equation} 
For an effective theory on a positive background and with an abelian gauge coupling, the condition \eqref{natcond} translates thus into
\begin{equation}
\label{Hg}
H \ll g M_P.
\end{equation}
Alternatively, as explained in \cite{Cribiori:2020use}, one can arrive at the same conclusion by remaining agnostic about \eqref{natcond} and \eqref{LIRH}, but by asking that corrections to the two-derivative effective action are suppressed, thus giving $H\ll \Lambda_{UV}$, and by invoking then the weak gravity conjecture.

If the weak gravity conjecture holds, the relation \eqref{Hg} is a consistency condition for any effective theory on a positive energy background and with an abelian gauge coupling. Clearly, it can be violated if the vacuum energy of the theory is such that
\begin{equation}
\label{Hquant}
H \sim g M_p,
\end{equation}
with no parameter which can be arbitrarily tuned entering the relation. The point is then whether or not this situation is indeed realised in concrete examples. As shown in \cite{Cribiori:2020use,DallAgata:2021nnr}, this is realised in de Sitter critical points (regardless of stability) of extended supergravity with a vanishing gravitino mass and with an unbroken abelian gauge group. In other words, \cite{Cribiori:2020use,DallAgata:2021nnr} showed that the Hubble scale in those extended supergravity models is quantised in terms of the UV cutoff given by the magnetic weak gravity conjecture. This implies that such models are not compatible with the condition \eqref{natcond}, or equivalently they are not protected against corrections, and thus cannot be consistent effective theories.\footnote{The same argument has been used in \cite{Cribiori:2020wch} to show that pure Fayet-Iliopoulos terms in $N=1$ supergravity are in the swampland, as it was already argued by \cite{Komargodski:2009pc} enforcing the absence of global symmetries. This is yet another confirmation of the fact that swampland conjectures are related to one another. } (They can still be consistent truncations.)
This can also be seen as a manifestation of the Dine-Seiberg problem \cite{Dine:1985he}, in the sense that the vacuum of the theory lies outside the region in which corrections are under control.

Before reviewing the analysis of \cite{Cribiori:2020use,DallAgata:2021nnr} showing that de Sitter critical points in $N=2$ supergravity with vanishing gravitino mass are of the type \eqref{Hquant}, let us mention possible loopholes in the argument above. First, it relies on the original formulation of the weak gravity conjecture in flat space. It might happen that corrections proportional to the spacetime curvature modify \eqref{mWGC} in such a way that the argument cannot be applied anymore. However, if the curvature is large, in order that we can safely assume gravity to be classical, it is reasonable to expect that corrections due to this effect are suppressed and can be neglected in first approximation. This seems to be in line with the analysis of \cite{Huang:2006hc} (see also \cite{Antoniadis:2020xso}). Another reason of concern is related to the condition \eqref{natcond}, which might not hold in an effective theory coupled to gravity, or at least not parametrically. In other words, it might be that gravity necessarily introduces an IR/UV mixing already at the level of the cutoff scales. This idea has been recently made precise in \cite{Castellano:2021mmx} (see \cite{Cohen:1998zx} for earlier work). It is a possibility we cannot exclude and whose effects on the argument would be interesting to investigate. However, as explained in \cite{Cribiori:2020use}, one can arrive at \eqref{Hg} even without starting from \eqref{natcond}, but just by requiring consistency of the two-derivatives effective action.

\subsection{The argument in $N=2$ supergravity}
\label{argN=2}

In this section, we review the general argument of \cite{DallAgata:2021nnr} showing that de Sitter critical points of $N=2$ supergravity with a vanishing gravitino mass (or parametrically lighter than the Hubble scale) are of the type \eqref{Hquant}, and thus cannot be used as consistent effective theories, according to the weak gravity conjecture. The argument is rather general and extends the analysis of \cite{Cribiori:2020use}. Besides asking for a vanishing gravitino mass, it requires the presence of an unbroken abelian gauge group in the vacuum, in order for the weak gravity conjecture to apply, but it covers both stable and unstable critical points. We follow the conventions of \cite{Ceresole:1995ca,Andrianopoli:1996cm}.

The sigma-trace of the gravitino mass matrix in $N=2$ supergravity is
\begin{equation}
S^x \equiv S_{AB} {(\sigma^{x})_C}^A \epsilon^{BC} = i \mathcal{P}^x_\Lambda L^\Lambda
\end{equation}
and it contributes to the scalar potential with a negative sign, $- 4S^x \bar S^x$. It is the only negative definite contribution to the vacuum energy and it vanishes if (sum over $x=1,2,3$ understood)
\begin{equation}
| \mathcal{P}^x_\Lambda L^\Lambda|^2=0.
\end{equation}
Under this assumption, the complete $N=2$ scalar potential,
\begin{equation}
\label{VN=2el}
\mathcal{V} = g_{i \bar\jmath} k^i_\Lambda k^{\bar \jmath}_\Sigma \bar L^\Lambda L^\Sigma + 4 h_{uv} k^u_\Lambda k^v_\Sigma \bar L^\Lambda L^\Sigma + (U^{\Lambda \Sigma} - 3 \bar L^\Lambda L^\Sigma) \mathcal{P}^x_\Lambda \mathcal{P}^x_\Sigma,
\end{equation}
reduces to
\begin{equation}
\begin{aligned}
\label{VN=2dS1}
\mathcal{V}_{dS} =-\frac12 ({\rm Im}\mathcal{N}^{-1})^{\Lambda \Sigma} (\mathcal{P}^x_\Lambda \mathcal{P}^x_\Sigma + \mathcal{P}^0_\Lambda \mathcal{P}^0_\Sigma) + 4 h_{uv} k^u_\Lambda k^v_\Sigma \bar L^\Lambda L^\Sigma ,
\end{aligned}
\end{equation}
where we used that the matrix $U^{\Lambda \Sigma}$ is defined as $U^{\Lambda \Sigma} = -\frac12 ({\rm Im}\mathcal{N}^{-1})^{\Lambda \Sigma} - \bar L^\Lambda L^\Sigma$ and we employed the special geometry relation $g_{i \bar\jmath} k^i_\Lambda k^{\bar \jmath} \bar L^\Lambda L^\Sigma = - \frac12 ({\rm Im} \mathcal{N}^{-1})^{\Lambda \Sigma}\mathcal{P}^0_\Lambda \mathcal{P}^0_\Sigma$, which can be derived from $\mathcal{P}^0_\Lambda L^\Lambda=0$.
Since the last term in \eqref{VN=2dS1} is positive definite, we can write
\begin{equation}
\mathcal{V}_{dS} \geq -\frac 12 {\rm Im}\mathcal{N}^{\Lambda \Sigma}  (\mathcal{P}^x_\Lambda \mathcal{P}^x_\Sigma + \mathcal{P}^0_\Lambda \mathcal{P}^0_\Sigma),
\end{equation}
and we recall that the matrix $({\rm Im} \mathcal{N}^{-1})^{\Lambda \Sigma}$ is negative definite. 
Now, we want to recast this expression in terms of the gravitino charge and gauge coupling, in order to apply the weak gravity conjecture. We follow a manifestly symplectic-covariant procedure, similar to that of \cite{Cribiori:2022trc}, which was performed for supersymmetric anti-de Sitter vacua. We introduce a SU(2) charge matrix 
\begin{equation}
2 {{\mathcal{Q}_\Lambda}_A}^B = \mathcal{P}^0_\Lambda \delta_A^B + {(\sigma_x)_A}^B \mathcal{P}_\Lambda^x, \qquad s.t. \quad {\rm Tr}\, \mathcal{Q}_\Lambda \mathcal{Q}_\Sigma = \frac12 \left(\mathcal{P}^0_\Lambda \mathcal{P}^0_\Sigma + \mathcal{P}^x_\Lambda \mathcal{P}^x_\Sigma\right).
\end{equation}
By employing the projectors ${{P^\parallel}^\Lambda}_\Sigma$,  ${{P^\perp}^\Lambda}_\Sigma$ defined as in \cite{Cribiori:2020use}, we can split $\mathcal{Q}_\Lambda = \mathcal{Q}_\Lambda^\parallel + \mathcal{Q}_\Lambda^\perp$, where
\begin{equation}
\label{Qparallel}
{{\mathcal{Q}_\Lambda^\parallel}_A}^B = \Theta_\Lambda {q_A}^B, \qquad  s.t. \quad {\rm Tr}\, Q^\parallel_\Lambda Q^\parallel_\Sigma = \Theta_\Lambda \Theta_\Sigma {\rm Tr}(q^2).
\end{equation}
Here, ${q_A}^B$ is the SU(2) charge entering the gravitino covariant derivative as 
\begin{equation}
D_\mu \psi_{\nu\,A} = \dots+ i \tilde A_\mu {q_A}^B \psi_{\nu \, B},
\end{equation}
where $\tilde A_\mu = \Theta_\Lambda A^\Lambda_\mu$ is the combination of vectors $A^\Lambda_\mu$ with coefficients $\Theta_\Lambda$ which is gauging the weak gravity U$(1)$ group. By canonically normalising the kinetic term of $\tilde A_\mu$, we can then read off the gauge coupling
\begin{equation}
\label{gcoup}
g^2 = - \Theta_\Lambda( {\rm Im} \mathcal{N}^{-1})^{\Lambda \Sigma} \Theta_\Sigma.
\end{equation}
The vacuum energy is thus rewritten as (using projectors' orthogonality)
\begin{equation}
\begin{aligned}
\mathcal{V}_{dS} &\geq - ({\rm Im}\mathcal{N}^{-1})^{\Lambda \Sigma} {\rm Tr} \, \mathcal{Q}_\Lambda \mathcal{Q}_\Sigma  \\
&=- ({\rm Im}\mathcal{N}^{-1})^{\Lambda \Sigma} ({\rm Tr} \, \mathcal{Q}^\parallel_\Lambda \mathcal{Q}^\parallel_\Sigma +{\rm Tr} \, \mathcal{Q}^\perp_\Lambda \mathcal{Q}^\perp_\Sigma ) \\
&\geq - ({\rm Im}\mathcal{N}^{-1})^{\Lambda \Sigma} {\rm Tr} \, \mathcal{Q}^\parallel_\Lambda \mathcal{Q}^\parallel_\Sigma .
\end{aligned}
\end{equation}
Employing \eqref{Qparallel} and \eqref{gcoup}, we get eventually
\begin{equation}
\mathcal{V}_{dS} \geq g^2 {\rm Tr} (q^2) \gtrsim {\rm Tr} (q^2) \Lambda^2_{UV},
\end{equation}
where in the last step we enforced the weak gravity conjecture. These vacua are of the form \eqref{Hquant} and thus cannot be consistent effective field theory descriptions. Notice that we are assuming some form of charge quantisation for the eigenvalues of ${q_A}^B$, but the precise details are not relevant for sake of the argument. The analogous proof for $N=8$ de Sitter critical points can be found in \cite{DallAgata:2021nnr}.

We see that asking for massless but charged gravitini on a positive energy background leads to an inconsistency according to the weak gravitino conjecture. This fact is also in agreement with the so called Festina Lente bound \cite{Montero:2019ekk,Montero:2021otb}, which proposes that $m^2\gtrsim \sqrt 6 g q H $ for every particle of mass $m$ and charge $q$. Thus, it rules out massless charged gravitini as well, albeit from a different perspective. Nevertheless, one expects all of the swampland considerations to be related to one another.

\subsubsection{Example: stable de Sitter from U$(1)_R\times$SO$(2,1)$ gauging}

Historically, the first examples of stable de Sitter vacua in $N=2$ supergravity have been proposed in \cite{Fre:2002pd}. One can check \cite{Cribiori:2020use} that the vacuum energy has precisely the form \eqref{Hquant}, but the gravitino mass vanishes at the critical point. Thus, the argument of the previous section applies and shows that those vacua are in tension with the weak gravity conjecture. As an illustrative example, below we review the simplest model of \cite{Fre:2002pd} from this perspective. 

The model has $n_V=3$ vector multiplets and no hyper multiplet. The $3+1$ vectors are gauging an SO$(2,1)\times$U$(1)_R$ isometry of the scalar manifold $\frac{{\rm SU}(1,1)}{{\rm U}(1)}\times \frac{{\rm SO}(2,2)}{{\rm SO}(2)\times {\rm SO}(2)}$. One can start from a basis with prepotential 
\begin{equation}
\mathcal{F} = -\frac12 \frac{X^1\left(\left(X^2\right)^2 - \left(X^3\right)^2\right)}{X^0}.
\end{equation}
In such a symplectic frame, the generators of SO$(2,1)$ are
\begin{equation}
T_0 =\scalebox{.65}{ $ \left(\begin{array}{cccccccc}
&&1&&&&&\\
&&&&&&-1&\\
\frac12&&&&&-1&&\\
&&&&&&&\\
&&&&&&-\frac12&\\
&&-\frac12&&&&&\\
&-\frac12&&&-1&&&\\
&&&&&&&
\end{array}
\right)$},
\quad
T_1 =\scalebox{.65}{ $ \left(\begin{array}{cccccccc}
&&-1&&&&&\\
&&&&&&1&\\
\frac12&&&&&1&&\\
&&&&&&&\\
&&&&&&-\frac12&\\
&&-\frac12&&&&&\\
&-\frac12&&&1&&&\\
&&&&&&&
\end{array}
\right)$},
\quad 
 T_2 = \scalebox{.65}{ $\left(\begin{array}{cccccccc}
-1&&&&&&&\\
&-1&&&&&&\\
&&&&&&&\\
&&&&&&&\\
&&&&1&&&\\
&&&&&1&&\\
&&&&&&&\\
&&&&&&&
\end{array}
\right),$}
\end{equation}
and we collect them into $T_\Lambda = (T_0, T_1, T_2, 0)$ for later convenience. When acting on $V^M=(X^\Lambda, F_\Lambda)$, the ${(T_\Lambda)^M}_N$ mix $X^\Lambda$ with $F_\Lambda$, i.e.~they generate so called non-perturbative symmetries, corresponding to gaugings which are not purely electric. To use the standard formula of the $N=2$ scalar potential \eqref{VN=2el}, we need to rotate the sections $V^M$ with an appropriate symplectic matrix,
\begin{equation}
V^M\to \tilde V^M = {S^M}_N V^N,
\end{equation}
where
 \begin{equation}
 {S^M}_N= \scalebox{.75}{ $\left(\begin{array}{cccccccc}
\frac12&&&&&1&&\\
-\frac12&&&&&1&&\\
&&1&&&&&\\
&&&\cos \delta&&&&\sin \delta\\
&-\frac12&&&1&&&\\
&-\frac12&&&-1&&&\\
&&&&&&1&\\
&&&-\sin\delta&&&&\cos \delta\\
\end{array}
\right)$}.
\end{equation}
Here, $\delta$ is a parameter defining the embedding of the U$(1)_R$ factor into Sp$(8,\mathbb{R})$ and it is historically known as de Roo-Wagemans angle \cite{deRoo:1985jh}. In the rotated basis the prepotential does not exist, but the new $X^\Lambda$ and $F_\Lambda$ are not mixed by the action of the SO$(2,1)$ generators. From this point on, we work in such basis for the symplectic sections, omitting the tilde for convenience. 

We now look at the scalar potential. Given that there are no hyper multiplets in the setup, the second term in the scalar potential \eqref{VN=2el} is vanishing, but the others are not. We analyse them separately. The last term (we denote it $\mathcal{V}_F$ below) is simpler since it is associated to the U$(1)_R$ gauging corresponding to a Fayet--Iliopoulos term. Indeed, we have a constant prepotential
\begin{equation}
\mathcal{P}^x_\Lambda= e_F \delta^{3x} \eta_{3\Lambda},
\end{equation}
with charge $e_F$, giving rise to the contribution
\begin{equation}
\mathcal{V}_F = e_F^2 \frac{(\cos \delta + x^1 \sin \delta )^2 + (y^1)^2\sin^2\delta }{2y^1},
\end{equation}
where we set $z^i = \frac{X^i}{X^0} = x^i-i y^i$.  Next, we concentrate on the first term in \eqref{VN=2el} (we denote it $\mathcal{V}_D$ below). This is non-vanishing since it corresponds to the gauging of SO$(2,1)$. The special K\"ahler prepotentials can be found with the general formula \cite{Andrianopoli:1996cm}\footnote{We are using the symplectic metric $\Omega_{MN} = \left(
\begin{array}{cc}
0&-1\\
1&0\\
\end{array}
\right).
$}
\begin{equation}
\mathcal{P}^0_\Lambda = e^K \bar V^M \Omega_{MN} {(T_\Lambda)^N}_P V^P,
\end{equation}
and in turn the killing vectors are computed as
\begin{equation}
k^i_{\Lambda} = i  g^{i\bar\jmath}\partial_{\bar \jmath} \mathcal{P}^0_\Lambda.
\end{equation}
One can check that the consistency conditions $L^\Lambda \mathcal{P}^0_\Lambda = 0 =k^i_\Lambda L^\Lambda$ are satisfied. The contribution to the scalar potential is then
\begin{equation}
\mathcal{V}_D = e_D^2 \frac{(x^3)^2 + (y^2)^2}{2y^1((y^2)^2-(y^3)^2)},
\end{equation}
where we inserted a charge $e_D$ to keep track of the SO(2,1) gauging. The total scalar potential is the sum of these two contributions
\begin{equation}
\mathcal{V} = \mathcal{V}_F + \mathcal{V}_D.
\end{equation}
It has a stable de Sitter vacuum at
\begin{equation}
x^1 = - \frac{1}{\tan \delta},\qquad y^1 =\frac{e_D}{e_F \sin \delta},\qquad x^3 = 0, \qquad y^3 =0,
\end{equation}
with vacuum energy
\begin{equation}
\label{VexdS}
\mathcal{V}_{dS} = e_D  e_F \sin \delta.
\end{equation}
Stability is explicitly verified in \cite{Fre:2002pd}.
Notice that the cosmological constant is strictly positive, since on the vacuum the K\"ahler potential is $K = - \log (4y^1 (y^2)^2) = -\log \left(\frac{4e_D^2 (y^2)^2}{\mathcal{V}_{dS}}\right)$. 

One can check that on the vacuum a subgroup U$(1)_R\times$U$(1)$ of the original gauge group is unbroken. Therefore, we can enforce the weak gravity conjecture with respect to the U$(1)_R$ factor. Furthermore, the gravitino mass is vanishing and thus the argument of section \ref{argN=2} applies.
It remains only to canonically normalise the kinetic term of the U$(1)_R$ vector and correctly identify the gauge coupling and charge. In the language of the general argument given previously, we have
\begin{equation}
2 {{\mathcal{Q}^\parallel_\Lambda}_A}^B = \left(0,0,0,e_F {(\sigma^3)_A}^B\right) \equiv 2 \Theta_\Lambda {q_A}^B,
\end{equation}
meaning
\begin{equation}
{q_A}^B = {(\sigma^3)_A}^B \mathrm{q},\qquad \Theta_\Lambda = \left(0,0,0,\frac{e_F}{2\mathrm{q}}\right),\qquad {\rm Tr}\,(q^2) = 2 \mathrm{q}^2,
\end{equation}
where $\mathrm{q}$ is the charge.
The kinetic matrix on the vacuum is (in the properly rotated symplectic frame)
\begin{equation}
({\rm Im}\mathcal{N}^{-1})^{\Lambda \Sigma} = -\sin \delta\,{\rm diag} \left( \frac{e_F}{e_D},\frac{e_F}{e_D},\frac{e_F}{e_D},\frac{e_D}{e_F}\right),
\end{equation}
and we thus find the gauge coupling
\begin{equation}
4g^2 \mathrm{q}^2 = e_D e_F \sin \delta.
\end{equation}
Eventually, we see that the vacuum energy \eqref{VexdS} has precisely the form \eqref{Hquant}. If the weak gravity conjecture holds, these vacua are not protected against corrections and thus are not within the controlled region of the effective description.

\section{The gravitino conjecture}
\label{sec_gravconj}

A crucial assumption in the discussion of the previous section is that the gravitino mass is vanishing (or very light with respect to the Hubble scale). It is then natural to wonder if the limit of parametrically small or even vanishing gravitino mass can be problematic for a given effective description more in general, regardless of the background. That this is indeed the case has been conjectured in \cite{Cribiori:2021gbf,Castellano:2021yye}, and it is the topic of the present section. We stress that we will be here concerned with the limit, while the argument of the previous section involved the actual evaluation of the gravitino mass at a specific (critical) point. Even if related, the two operations are in principle different.

\subsection{The statement and its motivation}

The gravitino conjecture \cite{Cribiori:2021gbf,Castellano:2021yye} states that the limit of vanishing gravitino mass, $m_{3/2}$, corresponds to a breakdown of the effective field theory description
\begin{equation}
m_{3/2}\to 0 \qquad \Rightarrow \qquad \text{swampland},
\end{equation}
as it is associated to the emergence of an infinite tower of states becoming light in the same limit. Some motivations behind this statement and based on  \cite{Cribiori:2020use,DallAgata:2021nnr} have already been reviewed in the previous section. Below, we would like to recall briefly some more. 

Clearly, for supersymmetric anti-de Sitter vacua, the statement is equivalent to the anti-de Sitter distance conjecture \cite{Lust:2019zwm} and thus, at least in this specific case, it shares the same motivation. However, the two conjectures crucially differ on other backgrounds and indeed they lead to differ predictions. Yet another motivation can be drawn from the idea of supersymmetric protection \cite{Palti:2020qlc}, which states that the superpotential of an $N=1$ theory cannot vanish, unless the theory is related in a specific way to a parent one with more supersymmetry. If the superpotential $W$ cannot vanish, the same applies to the gravitino mass, since in $N=1$ this is given by $m_{3/2}^2 = e^K W \bar W$. Instead, if the superpotential is exactly vanishing everywhere on the moduli space, the background is supersymmetric Minkowski, but this is not excluded by the conjecture, since the latter is really a statement about the limit (in the same way as the anti-de Sitter distance conjecture does not forbid supersymmetric Minkowski backgrounds).

Let us notice that unitarity alone seems not to be enough to motivate the gravitino conjecture on a positive background. Indeed, unitarity bounds for the gravitino have been studied in \cite{Deser:2001pe} and one can clearly see that there is no non-trivial unitarity bound on the gravitino mass if the cosmological constant is positive. It thus seems that the arguments reviewed in the previous section and the Festina Lente bound are capturing some more information on quantum gravity on a positive background beyond unitarity.

In a quasi-flat universe, the gravitino mass is a good approximation for the supersymmetry breaking scale,
\begin{equation}
M_{\text{susy}}^2 \simeq m_{3/2} M_P.
\end{equation}
The fact that supersymmetry has not been observed so far is thus very much compatible with the gravitino conjecture, which might point towards a scenario with supersymmetry breaking at high scale. This is naturally realised in string theory through a mechanism known as brane supersymmetry breaking \cite{Sugimoto:1999tx,Antoniadis:1999xk} (see \cite{Basile:2021vxh} for a recent review in a swampland perspective), in which the supersymmetry breaking scale is the string scale. Due to this fact, the effective descriptions of these models typically requires a non-linear realisation of supersymmetry in the low energy, which is not directly captured by the framework of \cite{Cribiori:2021gbf,Castellano:2021yye}. It would be interesting to extend the gravitino conjecture also to these setups.

Several examples in which the conjectures is checked explicitly can be found in \cite{Cribiori:2020use,DallAgata:2021nnr}. Furthermore, the new string construction proposed in \cite{Coudarchet:2021qwc} seem also to be compatible with it, since it is argued that supersymmetry is unavoidably (and softly) broken in the closed string sector. Below, we discuss and review some new and old examples, and we summarise various results. For concreteness, we will assume that the mass of the states becoming light is related to the gravitino mass as (in Planck units)
\begin{equation}
\label{m32n}
m_\text{tower} \sim m_{3/2}^n,
\end{equation}
with $n$ some model-dependent parameter. Clearly, this assumption might not be verified in general, but in the examples considered in \cite{Cribiori:2021gbf,Castellano:2021yye} does indeed hold. Furthermore, one can try to find bounds for $n$ or relate it to the parameters entering other swampland conjectures, as done in \cite{Castellano:2021yye}. It would be also interesting to check how the logarithmic corrections systematically proposed by \cite{Blumenhagen:2019vgj} can affect the analysis and the bounds on $n$.

\subsection{$N=1$ models}

As it is well-known, the structure of four-dimensional  $N=1$ supergravity with chiral and vector multiplets is completely fixed by three ingredients: a real gauge invariant function $G(z,\bar z)$, a holomorphic gauge kinetic function $f_{ab}(z)$, and a choice of holomorphic Killing vectors $k_a^i(z)$ generating the analytic isometries of the scalar manifold that are gauged by the vectors.  

Let us focus on the first ingredient, since it is closely related to the gravitino mass. Indeed, $G(z,\bar z)$ is defined as (assuming $W \neq 0$)
\begin{equation}
G(z,\bar z) = K + \log W \bar W,
\end{equation}
where $K$ and $W$ are the K\"ahler potential and superpotential respectively. This expression stems from the known fact that in supergravity, contrary to global supersymmetry, $K$ and $W$ are not functions but sections. In particular, they both vary under K\"ahler transformations. This means that they are not physical quantities: any statement regarding $K$ and $W$ separately depends crucially on a specific choice of K\"ahler frame. Instead, the only actual physical quantity to be considered is their K\"ahler invariant combination, $G(z,\bar z)$. The gravitino mass is then
\begin{equation}
m_{3/2}^2 = e^G
\end{equation}
and it is indeed K\"ahler-invariant.

We can give a geometric interpretation of the gravitino conjecture. The scalar manifold of $N=1$ supergravity is restricted by supersymmetry to be K\"ahler-Hodge (while in the global case it is just K\"ahler). This is a K\"ahler manifold $\mathcal{M}$ together with a holomorphic line bundle $\mathcal{L}\to \mathcal{M}$,  whose first Chern class is equal to the cohomology class of the K\"ahler form of $\mathcal{M}$, namely $c_1(\mathcal{L}) = [\mathcal{K}]$. Given a hermitean metric $h(z,\bar z)$ on the fiber, the first Chern class is $c_1(\mathcal{L}) = \frac{i}{2\pi} \bar \partial \partial \log h$. Setting it equal to the K\"ahler form $\mathcal{K} = \frac{i}{2\pi} \bar \partial \partial K$, one learns that for a K\"ahler-Hodge manifold the fiber metric is the exponential of the K\"ahler potential, $h = e^K$. The gravitino conjecture can then be geometrically rephrased as the statement that the limit of vanishing norm of a (non-vanishing) section of $\mathcal{L}$ leads to the breakdown of the effective theory. 
Indeed, consider a holomorphic section $W(z)$, which is in fact the superpotential of the $N=1$ theory. Its norm is defined as $W h \bar W = || W||^2$. Thanks to holomorphicity, one can write $ \frac{i}{2\pi} \bar \partial \partial \log h = \frac{i}{2\pi} \bar \partial \partial \log ||W||^2 $, identifying thus $G(z,\bar z) = K + \log W \bar W = \log ||W||^2 $. The gravitino mass is then
\begin{equation}
m_{3/2}^2 = e^G = || W||^2.
\end{equation}
The limit of vanishing gravitino mass corresponds to the limit of vanishing norm of the section $W(z)$. The gravitino conjecture implies that K\"ahler-Hodge manifolds compatible with quantum gravity are those in which this norm cannot be continuously sent to zero. Notice that special K\"ahler manifolds, parametrised by scalars of $N=2$ vector multiplets, are also K\"ahler-Hodge and thus this picture can be extended directly. One possible way in which the limit is obstructed consists in the existence of a lower positive bound for the gravitino mass. Simple models with this property are presented in section \ref{sec_modinvmod}. A possible refinement of the conjecture, inspired by the idea that domestic geometry is the natural framework underlying supergravity theories, as proposed in \cite{Cecotti:2021cvv}, consists in postulating that also the (dual) limit of infinite gravitino mass (or norm $||W||^2$) is pathological. This seems to be the case in a simple STU model reviewed in section \ref{stumodel}.

In order to test explicitly the conjecture, one should identify the tower of states becoming light in the limit of vanishing gravitino mass. Given that we are working within the framework of supergravity, a natural candidate for the first light tower are Kaluza-Klein states. Even if there might be subtleties in the precise identification of this scale (see e.g.~\cite{DeLuca:2021mcj} for recent work in this direction), we will then assume that their mass is parametrically related to the volume of the internal manifold. In Calabi-Yau or orbifold compactifications, the volume ({\rm Vol}) typically appears in the K\"ahler potential as
\begin{equation}
K(z,\bar z) = - \alpha \log {\rm Vol}  + K'(z, \bar z),
\end{equation}
where $\alpha$ is a model-dependent parameter, while the remaining part $K'$ will not play any role. In compactifications to four dimensions, one has that $m_{KK} \sim {\rm Vol}^{-\frac 23}$. We will also assume that the superpotential scales as
\begin{equation}
W \sim {\rm Vol}^{\frac \beta2},
\end{equation}
with $\beta$ a second model-dependent parameter. If $m_\text{tower} \sim m_{KK}$, we thus get a relation of the type \eqref{m32n}, with
\begin{equation}
n = \frac{4}{3(\alpha-\beta)}.
\end{equation}
This behavior can now be checked in concrete examples. As noted in \cite{Cribiori:2021gbf}, for heterotic compactifications to four-dimensional Minkowski spacetime one finds $n=\frac 43$, for type IIB GKP orientifolds one has $n=\frac23$ and for Scherk-Schwarz compactifications $n=4$. As discussed in \cite{Castellano:2021yye}, for general CY$_3$ orientifolds one finds a lower bound $n\geq \frac13$, while for F-theory flux compactifications on CY$_4$ one has $n\geq \frac 14$; however, for toroidal orientifolds the range seems to be further restricted to $\frac 23 \leq n \leq 1$.

\subsubsection{Modular invariant models}
\label{sec_modinvmod}

We would like to briefly review a class of models, not discussed in \cite{Cribiori:2021gbf,Castellano:2021yye}, in which the gravitino conjecture is satisfied by construction, since the gravitino mass admits a lower positive bound while varying over the moduli space. In these models, the interactions are fixed by asking that the action is left invariant by modular transformations acting on the scalar fields. Their construction in $N=1$ supergravity dates back to \cite{Ferrara:1989bc,Ferrara:1989qb}, while their string theory origin has been explored more in detail shortly after \cite{Font:1990nt,Font:1990gx,Cvetic:1991qm}. Recently, they have been revisited in a swampland context in \cite{Gonzalo:2018guu}.

In the simplest version, there is a single chiral multiplet $T$ transforming under SL$(2,\mathbb{Z})$ as
\begin{equation}
T \to \frac{aT + b}{cT + d}.
\end{equation}
The most generic $N=1$ Lagrangian invariant under this transformation is then given by
\begin{equation}
G(T,\bar T) = - 3 \log (-i (T-\bar T)) + \log W \bar W,
\end{equation}
with
\begin{equation}
W(T) = \frac{H(T)}{\eta(T)^6}.
\end{equation}
Here, $H(T)$ is a modular invariant holomorphic function and $\eta(T)$ the Dedekind function. It can be shown that, to avoid singularities in the fundamental domain, one has to choose
\begin{equation}
H(T) = (j(T) - 1728)^{\frac m2} j(T)^{\frac n3} P(j(T)),
\end{equation}
where $m$, $n$ are positive integers, and $P$ is a polynomial in the holomorphic Klein modular invariant form $j(T)$.

The origin of this superpotential is purely non-perturbative (gaugino condensation in heterotic compactifications). Therefore, the modulus $T$ which would be flat at the perturbative level is stabilised by non-perturbative corrections. The nature of these non-perturbative corrections is such that the decompactification limit, ${\rm Im} \, T \to \infty$, is prohibited, as the potential diverges in the same limit. This has to be contrasted with the standard behavior of perturbative potentials, which are typically vanishing at large distances. In other words, in these models a large distance limit is dynamically censored. This has consequences on the behavior of $m_{3/2}$ as a function of $T$:  the limit of vanishing gravitino mass is prohibited and the gravitino conjecture is automatically satisfied. 

As an illustration, let us consider the simplest superpotential which is realised in string compactifications,
\begin{equation}
W = \frac{1}{\eta(T)^6}.
\end{equation}
The gravitino mass is 
\begin{equation}
m_{3/2}^2 = e^G = \frac{1}{(2{\rm Im} \,T)^3 \eta(T)^6\bar \eta(\bar T)^6}
\end{equation}
and it can be easily checked that it is minimised at ${\rm Im} \, T=1$ at a strictly positive value. A plot is reported in the figure below.

\begin{figure}[h]
\begin{center}
\includegraphics[scale=.6]{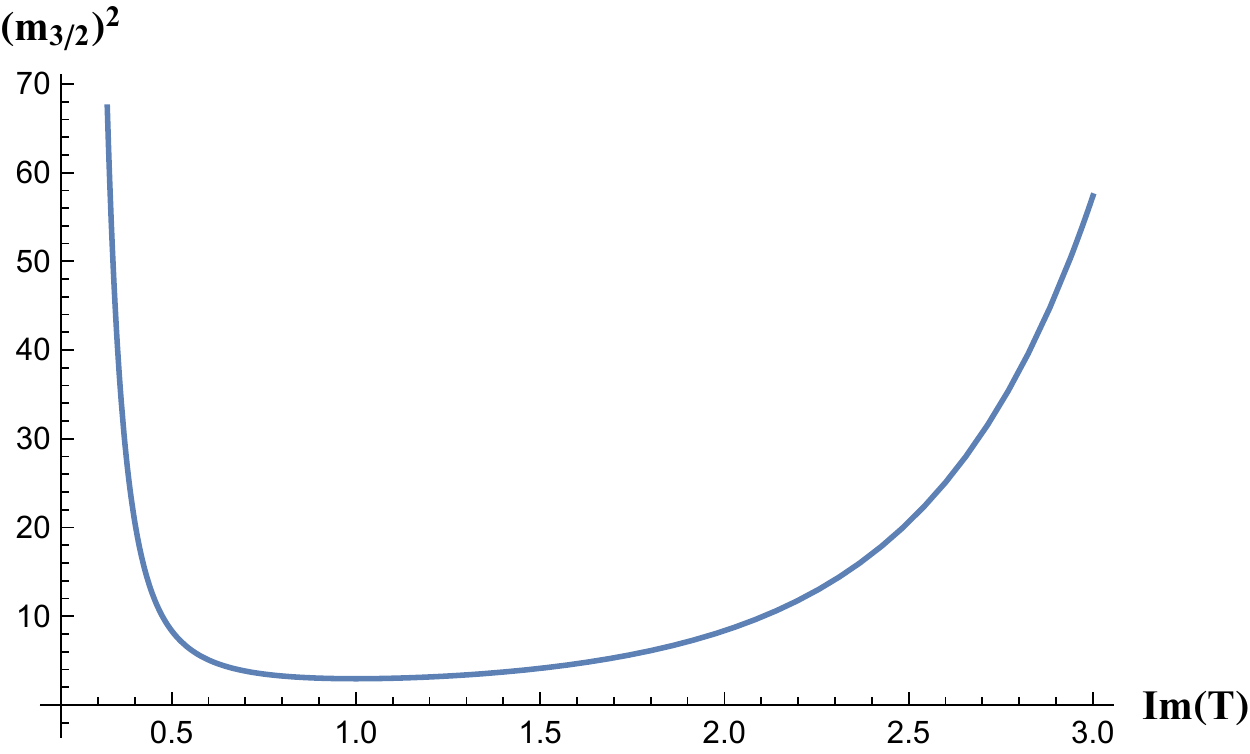}
\end{center}
\end{figure}

A natural generalisation would be to consider models with a non-trivial function $H(T)$ or with two chiral multiplets $S$ and $T$ \cite{Cvetic:1991qm}, described by $G(T,\bar T) = -\log(-i(S-\bar S))- 3 \log (-i (T-\bar T)) + \log W \bar W$, with $W =\frac{\Omega(S)H(T)}{\eta(T)^6}$.

\subsection{$N=2$ STU model}
\label{stumodel}

The so called STU model is an instructive and yet simple example in which to test the gravitino conjecture, as done in \cite{Cribiori:2021gbf}. It can arise from heterotic string compactification on $K3\times T^2$ and we refer e.g.~to \cite{deWit:1995dmj} for more details. The four-dimensional low energy theory is $N=2$ supergravity with $n_V=3$ vector multiplets, whose scalars are coordinates of the manifold $\left(\frac{SU(1,1)}{SU(1)}\right)^3$. The prepotential is 
\begin{equation}
\mathcal{F} = \frac{X^1 X^2 X^3}{X^0}.
\end{equation}

We consider the case of a U$(1)_R\subset$ SU$(2)_R$ gauging with a constant moment map, namely a Fayet--Iliopoulos term. For definiteness, we can choose it as
\begin{equation}
\mathcal{P}^x_\Lambda = q \delta^{3x}\eta_{0\Lambda}.
\end{equation}
In the presence of only Fayet--Iliopoulos terms, the $N=2$ scalar potential can be recast into an $N=1$ form \cite{Andrianopoli:1996cm}.  Indeed, from \eqref{VN=2el} with $k^i_\Lambda = 0 = k^u_\Lambda$ one finds\footnote{One has to use that $U^{\Lambda \Sigma} = g^{i\bar\jmath}f^\Lambda_i f^\Sigma_{\bar \jmath} = e^K g^{i\bar\jmath}D_i X^\Lambda \bar D_{\bar \jmath}\bar X^\Sigma$, where $g_{i\bar\jmath}$ is the special K\"ahler metric.}
\begin{equation}
\mathcal{V} = e^K \left(g^{i\bar \jmath}D_i W \bar D_{\bar \jmath}\bar W - 3 W \bar W\right),
\end{equation}
where we define the would be ``superpotential''
\begin{equation}
W = X^\Lambda \mathcal{P}^3_\Lambda. 
\end{equation}
The K\"ahler potential is
\begin{equation}
K = -\log( S T U)= - \log {\rm Vol},
\end{equation}
where $S = -2{\rm Im} z^1$, $T = -2{\rm Im} z^2$ and $U = -2{\rm Im} z^3$ and {\rm Vol} is the volume of the compact manifold. One can easily check that the scalar potential is identically vanishing and thus the model is no-scale. The gravitino mass matrix is 
\begin{equation}
S_{AB} = \frac i2 e^{\frac K2}W
\left(\begin{array}{cc}
0&1\\
1&0\\
\end{array}
\right).
\end{equation}
The mass of the gravitini is thus given by $m_{3/2}^2\simeq e^K W\bar W$, similarly to the $N=1$ case. Furthermore, the gauge coupling \eqref{gcoup} is
\begin{equation}
\label{gel}
g^2 = 2 e^{K} =2 {\rm Vol}^{-1}.
\end{equation}
We see that the limit $m_{3/2}^2 \to 0$ is realised by ${\rm Vol} \to \infty$. Interestingly, in the same limit the gauge coupling vanishes and a global symmetry is restored
\begin{equation}
{\rm Vol} \to \infty\qquad \Rightarrow \qquad m_{3/2} \to 0 \quad \Leftrightarrow \quad g \to 0.
\end{equation}
This would be problematic, since there should be no global symmetries in quantum gravity. This is a simple example in which the gravitino conjecture is closely related to the absence of global symmetries, and it would be interesting to explore further other relations of this kind within the web of swampland conjectures. In the limit of large volume, one can identity the tower of states becoming light as Kaluza-Klein modes. Notice that the T-dual limit of vanishing volume (very large gravitino mass) would also be problematic, since in this case winding modes would become light and the supergravity approximation would not hold anymore. Correspondingly, the magnetic dual of the gauge coupling \eqref{gel} would then vanish.

\section{Discussion: de Sitter and gravitino mass in extended supergravity}
\label{sec_fincomm}

The general argument of \cite{Cribiori:2020use,DallAgata:2021nnr}, reviewed in the section \ref{sec_WGCvsdS}, proves that de Sitter critical points of $N=2$ supergravity with vanishing gravitino mass have a cosmological constant of the order of the UV cutoff predicted by the magnetic weak gravity conjecture and thus cannot be consistent effective theories. 
The assumption of a vanishing gravitino mass might seem restrictive, but in \cite{Cribiori:2020use,DallAgata:2021nnr} it is found to hold true in almost all of the examples considered from the literature. In particular, in \cite{Cribiori:2020use} it is shown to occur in de Sitter critical points of minimal coupling models and cubic prepotential models with only vector multiplets, both for abelian and non-abelian gauging, while in \cite{DallAgata:2021nnr} explicit models with hyper multiplets are constructed. A model with massive gravitini in which the argument does not apply is the second de Sitter vacuum of \cite{DallAgata:2012plb}, which is unstable. A general strategy to construct de Sitter vacua with massive gravitini in $N=2$ supergravity has been proposed in \cite{Catino:2013syn}, but finding explicit models remains challenging \cite{DallAgata:2021nnr}.
Notice also that the argument requires an unbroken abelian gauge group in the vacuum, but in case there are only non-abelian groups it is still possible to formulate a weaker version, as explained in detail in \cite{DallAgata:2021nnr}. 

In \cite{DallAgata:2021nnr}, the argument has also been formulated in $N=8$ supergravity, and it should be possible to extend it to any $2<N<8$ theory. Assuming this step to be straightforward, we could conclude that, if the gravitino mass is vanishing or parametrically light compared to the Hubble scale, the resulting de Sitter critical points (regardless of stability) of extended supergravity are in the swampland due to the weak gravity conjecture. In turn, this would mean that $N=0,1$ supersymmetry at the Lagrangian level seem to be the most promising chances to obtain a consistent effective description for de Sitter critical points, if the gravitino mass is vanishing. Notice that this is compatible with the recent conjecture of \cite{Andriot:2022way} and has a nice parallel with the scale separation analysis of \cite{Cribiori:2022trc}. In particular, minimal supersymmetry in four dimensions is peculiar since it allows for the existence of a superpotential which might not be directly related to a gauge coupling. This possibility cannot occur in $N>2$, as it is most clearly illustrated in the example of the STU model reviewed in section \ref{stumodel}.

The analysis here presented was performed in four spacetime dimensions, but it would be interesting to extend it to higher dimensions. While we leave a detailed study for future work, we can provide an argument why de Sitter critical points of $d>4$, $N>0$ theories with vanishing gravitino mass suffer of the same problem discussed in section \ref{sec_WGCvsdS} and belong to the swampland, according to the weak gravity conjecture. First, one has to recall that in $d>4$ the supergravity scalar potential is always of order $\mathcal{O}(g^2)$, since it stems from a gauging procedure. Then, in any dimensions and for any number of supercharges, the structure of the scalar potential is fixed by the supersymmetric ward identity \cite{Cecotti:1984wn}, which has the schematic form
\begin{equation}
\mathcal{V} \sim \left(\delta (\text{spin-1/2})\right)^2 - \left(\delta (\text{spin-3/2})\right)^2.
\end{equation}
The precise numerical factors and indices entering this relation depend on $d$  (spacetime dimensions) and $N$ (number of supersymmetries), but crucially the relative negative sign between the two terms does not, since it corresponds to the gravitational contribution. Moreover, the second term in the relation above is precisely the gravitino mass, namely the supersymmetry transformation of the gravitino on the vacuum, with (Lorentz covariant) derivative and other maximal symmetry breaking terms turned off. Thus, if the gravitino mass is vanishing, the scalar potential is non-negative and of order $\mathcal{O}(g^2)$ for $d>4$. The argument of \cite{Cribiori:2020use,DallAgata:2021nnr}  then applies, provided one shows that there is no arbitrarily tunable parameter in the expression of the vacuum energy.

\acknowledgments
I would like to thank G.~Dall'Agata, F.~Farakos, D.~L\"ust and M.~Scalisi for discussions. This work is supported by the Alexander von Humboldt foundation.

\end{document}